\documentclass[twocolumn]{aastex63}

\usepackage{amsmath}
\usepackage{url}

\shorttitle{Post-outburst evolution of PSR J1119--6127}
\shortauthors{Blumer et al.}

\graphicspath{{./}{figures/}}

\begin{document}

\title{Back to quiescence: post-outburst evolution of the pulsar J1119--6127 and its wind nebula}

\correspondingauthor{Harsha Blumer}
\email{harsha.blumer@mail.wvu.edu}

\author{Harsha Blumer}
\affiliation{Department of Physics and Astronomy, West Virginia University, Morgantown, WV 26506, USA}
\affiliation{Center for Gravitational Waves and Cosmology, West Virginia University, Chestnut Ridge Research Building, Morgantown, WV 26505, USA}

\author{Samar Safi-Harb}
\affiliation{Department of Physics and Astronomy, University of Manitoba, Winnipeg, MB R3T 2N2, Canada}

\author{Alice Borghese}
\affiliation{Institute of Space Sciences (ICE, CSIC), Campus UAB, Carrer de Magrans s/n, 08193 Barcelona, Spain}
\affiliation{Institut d'Estudis Espacials de Catalunya (IEEC), 08034 Barcelona, Spain}

\author{Jonatan Mart{\'\i}n}
\affiliation{Institute of Space Sciences (ICE, CSIC), Campus UAB, Carrer de Magrans s/n, 08193 Barcelona, Spain}
\affiliation{Institut d'Estudis Espacials de Catalunya (IEEC), 08034 Barcelona, Spain}

\author{Maura A. McLaughlin}
\affiliation{Department of Physics and Astronomy, West Virginia University, Morgantown, WV 26506, USA}
\affiliation{Center for Gravitational Waves and Cosmology, West Virginia University, Chestnut Ridge Research Building, Morgantown, WV 26505, USA}

\author{Diego F. Torres}
\affiliation{Institute of Space Sciences (ICE, CSIC), Campus UAB, Carrer de Magrans s/n, 08193 Barcelona, Spain}
\affiliation{Institut d'Estudis Espacials de Catalunya (IEEC), 08034 Barcelona, Spain}
\affiliation{Instituci\'{o} Catalana de Recerca i Estudis Avan\c{c}ats (ICREA) Barcelona, Spain}

\author{George Younes}
\affiliation{Department of Physics, The George Washington University, Washington, DC 20052, USA}
\affiliation{Astronomy, Physics and Statistics Institute of Sciences (APSIS), The George Washington University, Washington, DC 20052, USA}

\begin{abstract}
We report on the analysis of a deep Chandra observation of the high-magnetic field pulsar (PSR) J1119--6127 and its compact pulsar wind nebula (PWN) taken in October 2019, three years after the source went into outburst. The 0.5--7 keV post-outburst (2019) spectrum of the pulsar is best described by a two-component blackbody plus powerlaw model with a temperature of 0.2$\pm$0.1~keV, photon index $\Gamma$=1.8$\pm$0.4 and X-ray luminosity of 1.9$_{-0.3}^{+0.3}\times$10$^{33}$~erg~s$^{-1}$, consistent with its pre-burst quiescent phase. We find that the pulsar has gone back to quiescence. The compact nebula shows a jet-like morphology elongated in the north-south direction, similar to the pre-burst phase. The post-outburst PWN spectrum is best fit by an absorbed powerlaw with a photon index $\Gamma$=2.3$\pm$0.5 and a flux of 3.2$^{+0.3}_{-0.2}$ $\times$10$^{-14}$~erg~cm$^{-2}$~s$^{-1}$ (0.5--7 keV). The PWN spectrum shows evidence of spectral softening in the post-outburst phase, with the pre-burst photon index $\Gamma$=1.2$\pm$0.4 changing to $\Gamma$=2.3$\pm$0.5 and pre-burst luminosity of 1.5$^{+0.5}_{-0.4}$$\times$10$^{32}$ erg~s$^{-1}$ changing to 2.7$_{-0.2}^{+0.3}$$\times$10$^{32}$ erg~s$^{-1}$ in the 0.5--7 keV band, suggesting magnetar outbursts can impact PWNe. The observed timescale for returning to quiescence, of just a few years, implies a rather fast  cooling process, and favors a scenario where J1119 is temporarily powered by magnetic energy following the magnetar outburst, in addition to its spin-down energy.
\end{abstract}


\section{Introduction}
\label{1}

Neutron stars, the evolutionary end-points of massive stars, are extremely compact remnants endowed with strong magnetic fields ($B$) ranging from $\sim$10$^{9}$--10$^{15}$~G. They are a diverse population, both in their observational and physical properties.  Rotation-powered pulsars (RPPs) and magnetars are two different manifestations of the neutron stars population. RPPs are mostly observed as rapidly-spinning, pulsating radio sources and are powered by their spin-down energy ($\dot{E}$).  Their periods $P$ span the range from 1~ms to 8 s and their $B$-field strengths range from $\sim$10$^{9}$--10$^{13}$~G.  Magnetars, on the other hand, are believed to be the strongest magnets in the universe with long $P$ $\sim$1--12~s and inferred surface dipole $B$-fields of 10$^{14}-10^{15}$~G, although some of them can have relatively weaker surface fields of $\gtrsim$10$^{12}$~G (see Kaspi \& Beloborodov 2017; Esposito et al. 2021 for reviews). They are characterized by intense episodes of X-ray and gamma-ray bursts, ranging from a few milliseconds duration to major month-long outbursts, followed by timing and spectral changes. Magnetar emission is believed to be powered by the decay of enormous internal magnetic fields, with indications that their magnetosphere is highly dynamic and characterized by a complex non-dipolar geometry (Thomson \& Duncan 1996). It is also expected that magnetars might be able to accelerate charged particles and produce extended, nebular outflows known as a magnetar wind nebulae (MWNe), which can provide important information on the composition and energetics of injected particles.  Many observational results in recent years, including the discovery of MWNe around the magnetar {\it Swift} J1834.9--0846 (Younes et al. 2016), have demonstrated that the above properties are not exclusive to magnetars, thereby bridging the gap between different neutron star classes. 
 
Among the RPPs, there is a peculiar class referred to as the high magnetic field pulsars (hereafter, high-$B$ pulsars) with inferred surface $B$-fields $\geq$4.4$\times$10$^{13}$~G, the quantum critical field limit. There are 7 high-$B$ pulsars, most of which have been discovered in the radio, except for the pulsar PSR J1846--0258 discovered in X-rays with no known radio counterpart. The high-$B$ pulsars further showed X-ray properties consistent with other lower-$B$ RPPs of comparable age as well as magnetars in quiescence, which led to the suggestion that these sources could be magnetars in disguise (e.g., Kaspi \& McLaughlin 2005, Safi-Harb \& Kumar 2008). Over the last two decades, different observational results proved that high-$B$ pulsars could indeed show magnetar-like bursts. The first such behavior was observed from PSR~J1846--0258 in 2006 (Gavriil et al. 2008; Kumar \& Safi-Harb 2008), followed by PSR J1119--6127 in 2016 (Archibald et al. 2016), which is the subject of study here. These results, along with the detection of bursts from other classes of neutron stars (e.g.,  central compact object (CCO) 1E 161348--5055 in the supernova remnant (SNR) RCW 103; Rea et al. 2016), further suggested that magnetars are more widely distributed than originally thought.

The high-$B$ field pulsar, PSR J1119--6127 (hereafter J1119), was discovered in the Parkes multibeam 1.4-GHz pulsar survey with a spin period $P$=408~ms and is associated with the supernova remnant (SNR) G292.2--0.5 (Camilo et al. 2000). The period, $P$, and the spin-down rate, $\dot{P}$=4$\times$10$^{-12}$~s~s$^{-1}$, imply a characteristic age $\tau_c\sim$1.9~kyr, a spin-down luminosity $\dot{E}$=2.3$\times$10$^{36}$~erg~s$^{-1}$ and a dipolar surface magnetic field $B$=4.1$\times$10$^{13}$~G. Observations performed with Chandra in 2002 detected the X-ray counterpart of J1119 and unveiled a faint, compact ($\sim$3\arcsec$\times$6\arcsec) wind nebula around it (Gonzalez \& Safi-Harb 2003). No radio PWN has yet been detected around the pulsar.  Only in 2004, when new Chandra observations were carried out, was it possible to study the PSR and the PWN independently (Safi-Harb \& Kumar 2008). The pulsar spectrum was well modelled by the combination of a blackbody (BB) of temperature $kT \sim$0.2~keV and a powerlaw (PL; characterized by the spectral index $\alpha$ or photon index $\Gamma$=$\alpha$+1) of $\Gamma \sim$ 2 to account for non-thermal emission above 3~keV. The PWN showed elongated jet-like features extending at least $\gtrsim$7$''$ north and south of the pulsar and its emission was described by a PL with $\Gamma$=1.1--1.4 (Safi-Harb \& Kumar 2008).  XMM-Newton observations of the source showed strong pulsations below 2.5 keV, with a pulsed fraction of 74$\pm$14\% (Gonzalez et al. 2005).  Furthermore, the pulsar has shown sporadic, or rotating radio transient-like behavior, preceded by large amplitude glitch-induced changes in the spin-down parameters in the radio wavelengths (Weltevrede et al. 2011).
  
Remarkably, on 2016 July 27, J1119 exhibited several short (0.02--0.04~s), energetic hard X-ray bursts detected by the Fermi Gamma-ray Burst Monitor and Neil Gehrels Swift Observatory (hereafter, Swift) Burst Alert Telescope which marked the onset of a magnetar-like outburst (G\"o\u{g}\"u\c{s} et al. 2016; Archibald et al. 2016). In the few days following these bursts, observations with the Swift X-ray Telescope (XRT) and NuSTAR showed that the unabsorbed 0.5--10 keV X-ray flux of J1119 had increased by a factor of $\sim$200, and strong X-ray pulsations were detected above 2.5 keV for the first time. The pulsar spectrum was dominated by a BB component with temperature $kT$ rising from 0.9~keV to 1.05~keV during the first two weeks of the outburst and a PL component with $\Gamma$=1.2 (Archibald et al. 2016, 2018). A hard X-ray component also suddenly appeared with emission extending at least up to $\sim$70 keV, a spectral behavior previously well established in many magnetars. The pulsar also underwent a contemporaneous spin-up glitch (Archibald et al. 2016). The radio emission was affected by the magnetar-like activity, initially becoming undetectable as a radio pulsar, and then returning with a steeper radio spectrum and a changed, multi-component pulse shape (Majid et al. 2017). The spin-down rate and X-ray flux increased by a factor of 5--10 before recovering toward the pre-burst rate (Archibald et al. 2018; Dai et al. 2018; Lin et al. 2018; Wang et al. 2020).  Thus, J1119 displayed a classic magnetar-like outburst, despite its normal appearance as a radio pulsar in the two decades since its discovery.

The pulsar's spectrum, obtained with Chandra three months after the outburst onset, was best described by a single PL model with $\Gamma$=2.0$\pm$0.2, softer than the value measured at the peak (Blumer at al. 2017). The pulsar luminosity was higher by a factor of $\sim$22 compared to its pre-outburst level, $L_{X,q} \sim$ 2$\times$10$^{33}$~erg~s$^{-1}$ (0.5--7~keV). The PWN's spectrum softened from a photon index of 1.2 to 2.2. The compact nebula also appeared brighter than its pre-burst state: the unabsorbed flux increased from 2.3$\times$10$^{-14}$~erg~cm$^{-2}$~s$^{-1}$  to 2.2$\times$10$^{-13}$~erg~cm$^{-2}$~s$^{-1}$ (0.5--7~keV). Moreover, we noticed a change in the PWN post-burst morphology, with a faint equatorial torus-like structure ($\sim$10\farcs0$\times$2\farcs5) along the southeast--northwest direction running perpendicular to a jet-like structure ($\sim$1\farcs5$\times$3\farcs5) southwest of the pulsar, while the pre-burst image showed a compact ($\sim$6$\arcsec$$\times$15$\arcsec$) PWN primarily extending along the north-south direction and visible only in the hard X-ray band (2.0--7.0 keV).

In this paper, we report on a detailed study of the post-outburst evolution of J1119 and its associated nebula with Chandra, along with a reanalysis of all archived Chandra observations for consistency and for studying the evolution of the pulsar and its PWN. J1119 is associated with the SNR G292.2--0.5 at a distance of 8.4~kpc (Caswell et al. 2004); therefore we scale all derived quantities in units of $d_{8.4}$=$D/8.4$~kpc. The paper is organized as follows: Section 2 describes the observation and data reduction. We present the details of the X-ray imaging analysis in Section 3 and spectral fitting in Section 4. The results are discussed in Section 5, including modelling the observations in light of the recently introduced model discussing the effect of magnetar bursts on the nebula (Mart{\'\i}n et al. 2020). Finally, we summarize our findings in Section 6. 

\begin{figure*}[ht]
\includegraphics[width=\textwidth]{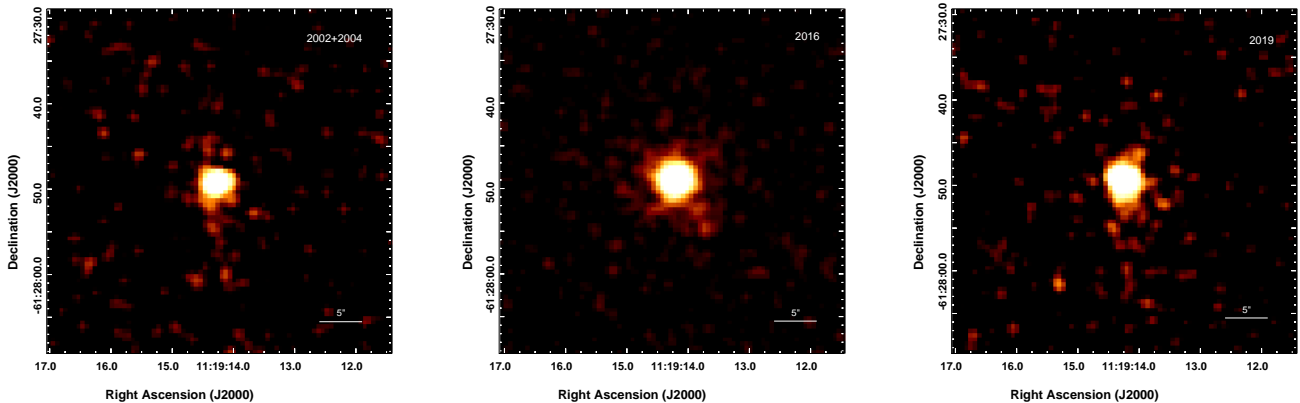}
\caption{Broadband (0.5--7 keV) images of PSR~J1119--6127 and its surrounding PWN (in logarithmic scale) for the pre-burst (left), burst (middle), and post-outburst (right) Chandra data. The pre-burst image was obtained by combining the 2002 and 2004 observations while the post-outburst image was made by combining all the four observations in 2019. The images are exposure-corrected and smoothed using a Gaussian function of radius 2 pixels. North is up and East is to the left. }
\end{figure*}

\section{Observation and Data Reduction}
\label{2}

J1119 was previously observed with the Chandra X-ray observatory in 2002 and 2004 during its quiescence, and in 2016, three months after the onset of magnetar-like outbursts. We obtained a deep new observation of J1119 on 2019, split into four pointings between October 18--26, three years after the source went into outburst. All the observations were positioned at the aimpoint of the back-illuminated S3 chip of the Advanced CCD Imaging Spectrometer (ACIS). The standard processing of the data was performed with the Chandra Interactive Analysis of Observations software package (CIAO version 4.12; Fruscione et al. 2006) and the calibration files in the CALDB database (version 4.9.1). The event files were reprocessed (from level 1 to level 2) to remove pixel randomization and to correct for CCD charge transfer efficiencies.  An examination of the background light curves did not show any strong flares. We have also reprocessed all previous observations of J1119 obtained with Chandra, similarly.  The resulting effective exposure times for the observations are given in Table~1. 

\begin{table}
\center
\caption{Log of Chandra observations of PSR J1119--6127}
\begin{tabular}{l l l}
\hline\hline Obsid & Date & Exposure (ks) \\
\hline
2833 & 31 March 2002 & 56.79\\
4676 & 31 October 2004 & 60.54\\
6153 & 2 November 2004 & 18.90\\
19690 & 27 October 2016 & 55.50\\
22422 & 18 October 2019 & 52.56\\
22877 & 21 October 2019 & 44.47\\
22883 & 24 October 2019 & 17.82\\
22884 & 26 October 2019 & 19.80\\
\hline 
\end{tabular}
\tablecomments{Exposure represents the effective exposure times obtained after following the CIAO routines.}

\end{table}

\section{Imaging analysis}
\label{3.1}

Figure~1 shows the 0.5--7 keV image of J1119 and its compact nebula for the three different epochs, marked as pre-burst (left), burst (middle), and post-outburst (right). The pre-burst, burst, and post-outburst epochs refer to the Chandra observations taken in 2002+2004, 2016, and 2019, respectively.  The image is centered on the pulsar coordinates at $\alpha_{J2000}$=11$^{h}$19$^{m}$14$^{s}.260$ and $\delta_{J2000}$=$-$61$^{\circ}$27$\arcmin$49$\arcsec$.30. The images have been exposure-corrected using the CIAO task \textit{fluximage} with a binsize of 1 pixel and smoothed using a Gaussian function of radius 2 pixels. The pre-burst image was made by combining the 2002 and 2004 observations, and shows a compact nebula of size 6$\arcsec$$\times$15$\arcsec$ in the north-south direction (Safi-Harb \& Kumar 2008). The 2016 burst image clearly shows a brighter nebula and small-scale fine structures ($\sim$10$\arcsec$) around the pulsar (see Blumer et al. 2017 for details). The post-outburst image was made by combining all the four observations in 2019 and features a much fainter PWN, with elongated jet-like features along the north-south direction, similar to its pre-burst phase. 

To further investigate the PWN morphological differences between the three epochs, we extracted the radial profile of the observed (0.5--7~keV) surface brightness up to a radial distance of 15$\arcsec$ from J1119, using as a background an annulus with radii 30$\arcsec$ and 40$\arcsec$. We compared the observed radial profiles with the point-spread functions (PSFs),  simulated with the Chandra Ray Tracer (ChaRT; Carter et al. 2003) and MARX package (Davis et al. 2012) adopting the best-fit pulsar spectrum and for an {\sc AspectBlur} parameter of 0.28 (see Figure~2). Figure~3 shows the surface brightness profiles for J1119 for the three epochs, where the post-outburst profiles are consistent with the pre-burst phase and excess is seen at radial distances $\gtrsim$2$\arcsec$.

\begin{figure*}[ht]
\includegraphics[width=\textwidth]{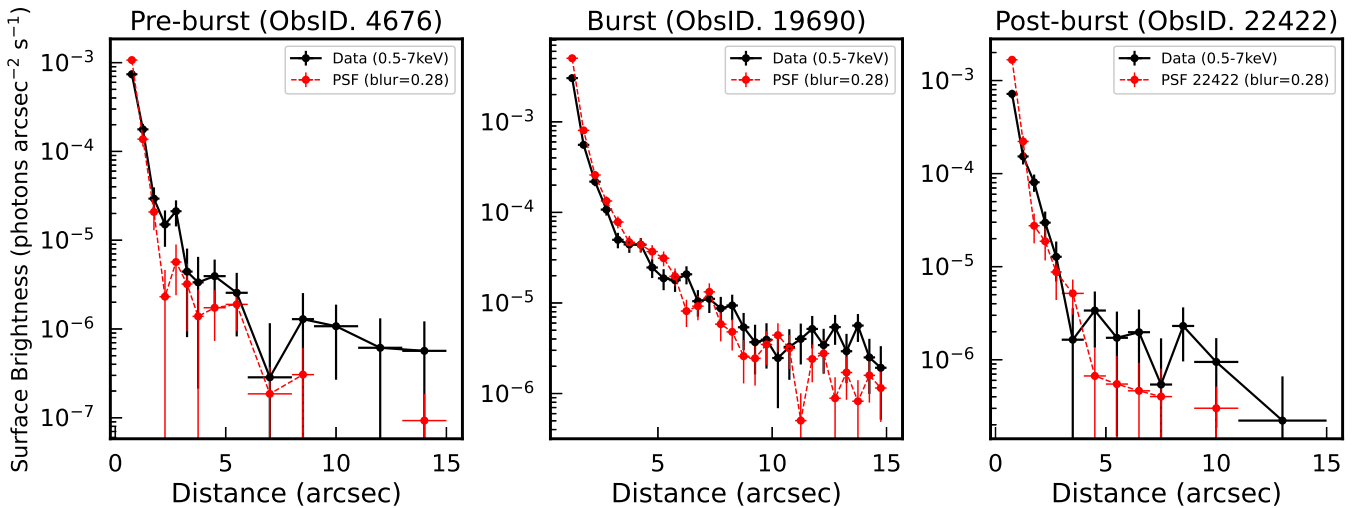}
\caption{Left: Observed surface brightness profiles of J1119 (in red) for the pre-burst, burst, and post-outburst epochs up to a radial distance of 15$\arcsec$ from J1119 in the 0.5--7\,keV energy range. The simulated PSFs are shown in black for an aspect blur value of 0.28. See Section 3 for details.}
\end{figure*}

\begin{table*}[ht]
\begin{center}
\caption{Spectral fits to the PSR J1119--6127 and its PWN}
\begin{tabular}{l l l l l l l}
\hline\hline
Parameter & \multicolumn{3}{c}{Pulsar} &  \multicolumn{3}{c}{PWN} \\
\cline{2-7}
 & 2002+2004 & 2016 & 2019 & 2002+2004 & 2016 & 2019 \\
& BB+PL & PL & BB+PL & PL & PL & PL\\

\hline

$N_{H}$ ($10^{22}$ cm$^{-2}$) &  & 2.0$\pm$0.5 &  & & 2.2$\pm$0.4 \\

kT (keV) &  0.2$\pm$0.1 & \nodata & 0.2$\pm$0.1  & \nodata  &  \nodata & \nodata\\

$\Gamma$ & 1.7$\pm$0.5 & 2.0$\pm$0.2 & 1.8$\pm$0.4  & 1.2$\pm$0.4 & 2.0$\pm$0.6 & 2.3$\pm$0.5 \\

$F_{unabs}$ (BB)$^a$ & 1.3$_{-0.2}^{+0.3}\times10^{-13}$ & \nodata & 1.7$^{+0.3}_{-0.2}\times10^{-13}$ &  \nodata  &  \nodata & \nodata \\

$F_{unabs}$ (PL)$^a$ & 5.5$_{-0.5}^{+0.4}\times10^{-14}$ & 5.3$^{+0.4}_{-0.5}$$\times$10$^{-12}$ & 6.2$_{-0.4}^{+0.3}\times10^{-14}$ & 1.8$_{-0.5}^{+0.6}$$\times10^{-14}$ & 2.0$_{-0.2}^{+0.2}$$\times10^{-13}$ & 3.2$_{-0.2}^{+0.3}$$\times10^{-14}$  \\

$L_{X}$$^b$ & 1.6$_{-0.4}^{+0.3}\times$10$^{33}$ & 4.5$_{-0.4}^{+0.3}\times$10$^{34}$ & 1.9$_{-0.3}^{+0.3}\times$10$^{33}$ & 1.5$_{-0.4}^{+0.5}\times$10$^{32}$ & 1.7$_{-0.2}^{+0.2}\times$10$^{33}$ & 2.7$_{-0.2}^{+0.3}\times$10$^{32}$\\

$L_{X}$/$\dot{E}$ & 0.001 & 0.02 & 0.001 & 0.65$\times$10$^{-4}$ & 7.0$\times$10$^{-4}$ & 1.7$\times$10$^{-4}$ \\

Count rate$^c$ & (5.3$\pm$0.2)$\times$10$^{-3}$ & (1.1$\pm$0.0)$\times$10$^{-1}$ & (5.7$\pm$0.2)$\times$10$^{-3}$ & (8.3$\pm$2.3)$\times$10$^{-4}$ & (6.9$\pm$0.4)$\times$10$^{-3}$  & (1.1$\pm$0.1)$\times$10$^{-3}$  \\

\hline
\end{tabular}
\end{center}
\tablecomments{The 2002+2004, 2016, and 2019 epochs refer to the \textit{Chandra} observations taken during the pre-burst, burst, and post-outburst phase, respectively. $N_H$ is tied together for all 3 epochs for the pulsar and PWNe. Errors are at 90\% confidence level. \\
$^a$ Unabsorbed flux (0.5--7 keV) in units of erg cm$^{-2}$ s$^{-1}$. \\
$^b$ X-ray luminosity (0.5--7 keV) in units of erg s$^{-1}$ assuming isotropic emission at a distance of 8.4~kpc. \\
$^c$ Background subtracted count rates (0.5--7 keV) in units of counts s$^{-1}$.
 }
\end{table*}

\section{Spectral analysis}
\label{4}

The spectra were extracted for the source and background regions and the corresponding response files were created using {\it specextract}, and analyzed with the XSPEC (v12.10.1f; Arnaud 1996) fitting package. The contributions from background point sources were removed prior to the extraction of spectra.  We then combined the spectra created for each ObsID in the pre-burst (2002+2004) and post-outburst (2019) phase (see Table~1) using the CIAO tool {\it combine\_spectra}  to make a single data set for each epoch, which has the advantage of increasing the signal-to-noise ratio for spectral analysis. The resulting combined spectra were rebinned to have at least 10 counts per spectral bin and errors are at 90\% confidence level.  The spectral analysis was restricted to the 0.5--7 keV band, where the pulsar and PWN signal-to-noise ratio was higher. We used the $tbabs$ model (Wilms et al. 2000) to describe photoelectric absorption by the interstellar medium.

The spectrum of J1119 was extracted from a 1\farcs5 radius circular region centred on the source, which encompasses more than $\sim$90\% of the encircled energy for a point source observed on-axis with \textit{Chandra}\footnote{http://cxc.harvard.edu/proposer/POG/html/chap6.html} at 1.49 keV.  The background was chosen from an annular ring of 3$\arcsec$--5$\arcsec$ centered on the source. We also estimated the impact of photon pileup in all the Chandra observations using WebPIMMS (version 4.10) and \textit{jdpileup} model of the \textit{Chandra} spectral fitting software \textit{Sherpa} (Freeman et al. 2001; Doe et al. 2007; Burke et al. 2020) convolved with an absorbed PL or BB model to the pulsar spectrum. The 2016 data were affected by 10\% pileup, while all other data had negligible pileup. Therefore, we included a pileup model (Davis et al. 2001) as implemented in XSPEC for the 2016 data during spectral fitting. 

The pulsar spectra for all three epochs were simultaneously fit with different one- and two-component models,  The column density $N_H$ was tied between the epochs, leaving all other parameters to vary during the fit.  As found by Safi-Harb \& Kumar (2008), a single component BB model did not yield a good fit to the spectrum of the pulsar in quiescence. We confirm that a hard PL component in addition to the BB component was needed to fit both the pre-burst (2002+2004) and the most recently acquired (2019) spectra. The 2016 (burst) data were however adequately fitted by a single component PL model, as illustrated in Blumer et al. (2017).  The results of the spectral fits are shown in Table~2.  We obtained consistent results when fitting the spectra individually.  Figure~3 (left) shows the best-fit pre-burst (BB+PL; black), burst (PL; red), and post-outburst (BB+PL; green) pulsar spectra.

\begin{figure*}[ht]
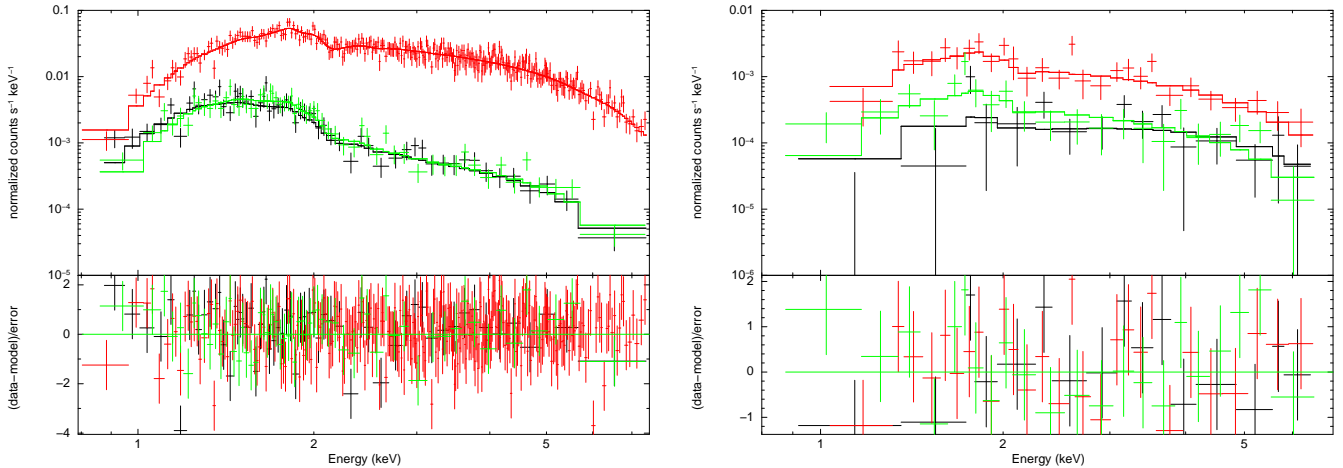

\includegraphics[angle=-90, width=0.5\textwidth]{fig3a.eps}
\includegraphics[angle=-90, width=0.5\textwidth]{fig3b.eps}
\caption{{\it Left}: X-ray spectra of PSR~J1119--6127. The pre-burst (black) and post-outburst (green) data are best fit by a BB+PL model while the burst data (red) are fit by a PL model. {\it Right}: X-ray spectra of the compact PWN fitted with a PL model. The topmost, middle, and bottom spectra represent the pre-burst (black), burst (red), and post-outburst (green), respectively.  The lower panel shows the ratio of data to plotted model. }
\end{figure*}

To extract the photons from the PWN, we followed the same procedure as in Blumer et al. (2017) and selected an annular region of 2$\arcsec$ to 10$\arcsec$ radii to ensure that all small-scale features were included. The background was extracted from a nearby source-free elliptical region. To model the PWN emission and investigate any spectral variations between the epochs, we fit the three spectra simultaneously with an absorbed PL model, tying only the $N_H$ among the data sets and allowing the PL photon indices and normalizations to vary. The absorption $N_H$ was found to be (2.2$\pm$0.4)$\times$10$^{22}$~cm$^{-2}$, consistent within errors to that obtained for the pulsar. The PWN spectral parameters are listed in Table~1. The spectral fits were also explored with different annular and elliptical backgrounds and binning, and the spectral parameters agree within the errors to those reported in Table 1. Figure~3 (right) shows the best-fit PL model for the PWN.

\section{Discussion}
\label{4}

Thanks to the prompt response and dedicated follow-up programs of Chandra and other X-ray satellites, the past decade has seen a great success in detecting many magnetar outbursts and studying their post-outburst emission mechanisms. The spectacular angular resolution of Chandra has allowed us to collect an unprecedented dataset for the high-$B$ pulsar J1119 and its compact nebula covering the period from 2002 to 2019. This has enabled us to accurately characterize the behavior of the source over a long time span of $\sim$17 yrs. We discuss the implications of our findings below.

The persistent soft X-ray spectrum of magnetars usually comprises a thermal (BB temperature $kT$$\sim$0.3--0.6 keV) and a nonthermal (photon index $\Gamma$$\sim$2--4) component. According to the twisted-magnetosphere model of magnetars, the thermal emission originates from heating within the star due to the decay of strong internal magnetic fields (Thompson et al. 2002). Magnetospheric currents, which circulate in localized bundles of magnetic field lines, scatter the thermal surface photons to higher energies. These currents provide a source of surface heating in the form of a return current. The flux increase that accompanies a magnetar outburst is suggested to be due to a rapid heating originating from magnetospheric, internal, or crustal reconfiguration of the neutron star (Thompson et al. 2002; Beloborodov 2009). The twisted magnetosphere does not remain static and gradually untwists, dissipating magnetic energy and producing radiation. 

The quiescent spectrum of J1119 is well described by a BB of temperature $kT$=0.2$\pm$0.1~keV and PL of index $\Gamma$=1.7$\pm$0.5, which made this pulsar the youngest with detected thermal emission (see also Safi-Harb \& Kumar 2008). Following the X-ray outburst in 2016, its spectrum changed to mainly nonthermal in nature, with an index $\Gamma$=2.0$\pm$0.2 (Table~2). The spectra of magnetars generically show a hardening at the time of outburst, and then slowly soften as the flux relaxes back to quiescence over months to years (Kaspi \& Beloborodov 2017). This hardening at outburst is generally seen as an increase in $kT$ and a decrease in the photon index for a BB+PL model. For J1119, although a second component was not needed for the 2016 burst phase, a BB+PL model fit shows evidence of hardening with $N_H$=2.2$\pm$0.4~cm$^{-2}$, $kT$=0.4$\pm$0.2, $\Gamma$=1.5$\pm$0.4, and the PL flux dominating the emission (see Blumer et al. 2017). The presence of this dominant nonthermal component is suggestive of a large density of magnetospheric particles, which boost thermal photons emitted from the surface via resonant Compton scattering, providing the powerlaw component. 

The 2019 post-outburst spectrum of J1119 yielded a BB temperature $kT$=0.2$\pm$0.1~keV, photon index $\Gamma$=1.8$\pm$0.4 and an unabsorbed total flux of 2.3$_{-0.5}^{+0.4}\times$10$^{-13}$~erg~cm$^{-2}$~s$^{-1}$, with the thermal flux dominating the X-ray emission.  Assuming isotropic emission, J1119's post-outburst X-ray luminosity is $L_{X}$=4$\pi$$d_{8.4}^2$$F_{unabs}$=1.9$_{-0.3}^{+0.3}$$\times$10$^{33}$~$d_{8.4}^2$~erg~s$^{-1}$, implying an efficiency $\eta_{X}$=$L_{X}/\dot{E}$$\sim$0.001~$d_{8.4}^2$ in the 0.5--7 keV energy range. These values are consistent with those obtained for its pre-burst data, implying that the pulsar has mostly returned to its quiescence, and in line with the outburst decay trend observed in other magnetars and high-$B$ pulsars. Recent studies show that most magnetars had reached full quiescence within $\sim$1000 days after the outburst onset (Coti-Zelati et al. 2018).  The magnetospheric untwisting model by Beloborodov (2009) also makes predictions regarding the post-outburst flux and spectral evolution of magnetars. In this framework, the post-outburst magnetosphere has been twisted due to crustal motions originating from stresses induced by the strong internal magnetic field. The twist is carried by a bundle of current-carrying field lines, the so-called j-bundle, which is anchored on a footprint in the crust.  The magnetic twist energy is dissipated over time in the form of thermal radiation during relaxation and untwisting.  The energy released in the j-bundle can further power nonthermal magnetospheric emission near the dipole axis, thus feeding a PL component at late stage of the outburst (Beloborodov 2009).

The compact PWN around J1119 in the post-outburst observations shows a morphology similar to its pre-burst phase, with elongated jet-like features in the north-south direction (see Figures~1 and 2). Its spectrum is well described by a PL with photon index $\Gamma$=2.3$\pm$0.5, unabsorbed flux of 3.2$_{-0.2}^{+0.3}\times$10$^{-14}$~erg~cm$^{-2}$~s$^{-1}$, and X-ray luminosity of 2.7$_{-0.2}^{+0.3}\times$10$^{32}$~$d_{8.4}^2$~erg~s$^{-1}$ in the 0.5--7 keV band. The inferred X-ray efficiency is $\eta_X$=1.7$\times$10$^{-4}$, which is consistent with the typical values of $\sim$10$^{-5}$ to 10$^{-1}$ observed in other rotation-powered pulsars with PWNe (Kargaltsev \& Pavlov 2008). During the 2016 observation, the PWN had showed evidence of brightening and a softer photon index (2.0$\pm$0.6), likely combined with a dust scattering halo (Blumer et al. 2017). Interestingly, the 2019 post-outburst observations also show a softer index for the PWN with a decrease in the flux, comparable to its pre-burst quiescent values (Table~2). To further quantify the spectral fit results, we estimated the number of soft and hard counts for the three epochs. For the pre-burst, burst, and post-outburst phase, the soft (0.5--2 keV) photon counts are 10$\pm$9, 85$\pm$11, and 62$\pm$10, while the hard (2--7 keV) photon counts are 86$\pm$14, 164$\pm$14, and 83$\pm$13, respectively. There is an increase in the soft photons for the post-outburst epoch with respect to the pre-outburst phase, while the number of hard photons is consistent between the two epochs. These results further confirm the spectral softness observed in the PWN, which is likely impacted by the outburst.

Whether magnetars and high-$B$ pulsars power PWNe or magnetar wind nebulae (MWNe) is an open and interesting question. Magnetars are believed to produce relativistic particle outflows, either steady or during bursting episodes (Harding et. al. 1999). Clear evidence for temporary magnetar outflows has also been seen in the form of transient extended radio emission following two giant flares (Gaensler et al. 2005). However, MWNe are considered long-lived and result from continuous particle outflow even when the magnetar is in quiescence. The most compelling case of a MWN is an asymmetrical X-ray structure around the magnetar Swift J1834.9--0846 (Younes et al. 2016), where the extended emission remained fairly constant in flux and spectral shape across 9 years of observations. In the X-ray range, where diffuse emission has been seen around several magnetars, the identification of a MWN is complicated by the formation of a dust scattering halo accompanying a magnetar burst due to the large interstellar absorption and high X-ray luminosity of these sources.  J1119 had not shown any magnetar-like bursts since 2016 and hence, we do not expect to see any scattering halo in the new Chandra observations.  Small scale variabilities are expected in PWNe, however, the photon index of J1119's nebula has remained softer even after 3 yrs of magnetar-like activity.

Changes in flux are, however, noticeable beyond doubt. Three months after the burst, the PWN was $\sim$10 times brighter than in its pre-outburst state. However three years after the burst, the PWN came back to its pre-burst flux level. This raises the question whether the rise and decrease of the flux level at the PWN before and after the outburst of 2016 is indeed physically connected with it or not. We cannot rule out the idea that the PWN X-ray flux increase could be the result of a process of release of energy, magnetic field, and particles that started before the detection of the pulsar burst itself. This idea, although possible in principle, is the less testable (and the less appealing) since it implies a total disconnection between the rise and decay times of the magnetar burst observed in 2016 and the PWN phenomenology following to it. In what follows, then, we consider that the most natural scenario is that the recent PWN phenomenology is indeed related to the influence of the 2016 burst.

In recent work, Mart{\'\i}n et al. (2020) explored possible effects of magnetar bursts on the radio, X-ray, and gamma-ray fluxes of PWNe assuming either that the burst injects electron-positron pairs or powers the magnetic field. Similar to gamma-ray bursts (GRBs), they considered that the magnetar flare could be associated with an outflow carrying kinetic and magnetic energy that collides with the PWN, driving a forward shock and accelerating electrons to higher energies. The magnetic field in the forward shock region would also be enhanced with respect to the original PWN field, e.g., due to shock compression and/or Weibel instability in the shock downstream. Mart{\'\i}n et al. (2020) considered that a significant amount of relativistic particles is injected in the PWN as a result of a magnetar burst -- happening roughly instantaneously in comparison to the dynamical timescales of the PWN. For instance, in an injection of particles during 1 second with a total energy $E_{out}$ of 10$^{45}$, 10$^{46}$ and 10$^{47}$ erg, when particles reach the termination shock, the luminosity is increased by factors of 10--20 in X-rays in the most extreme cases. From the perspective of the flux increase, this could work. However, the loss timescales for particles are too large in comparison to a decay of a few years, as in the case of J1119. In fact, we expect such an enhanced luminosity to remain high for several kyr. The same happens if the energy of the burst is considered to power the magnetic field, which then decays via adiabatic losses with the PWN expansion velocity (see Eq. 15 of Mart{\'\i}n et al. 2020).

Shorter decay times can be approached by considering that the evolution equation of the magnetic field energy of the perturbation $E_{\Delta B}$ is
\begin{equation}
\frac{d E_{\Delta B}}{dt}=\eta' L'(t)-\frac{E_{\Delta B}}{R} \frac{dR}{dt}
\end{equation}
where $R$ is the radius of the perturbation wave which we take as $R \simeq c t$, and we assume that the injection has the form
\begin{equation}
L'(t)=L'_0 e^{-\frac{t-t_0}{t_{decay}}}
\end{equation}
where $t_0$ is the time when the injection starts, $t_{decay}$ the decay injection timescale and $L'_0$ the initial injection luminosity. If the total energy injected is $E_{out}$, a fraction  $\eta'$ of such energy will sustain the magnetic perturbation. To determine the value of $L'_0$ we can use energy conservation
\begin{equation}
E_{out}=\int_{t_0}^\infty L'_0 e^{-\frac{t-t_0}{t_{decay}}} \mathrm{d}t
\end{equation}
resulting in $L'_0=\eta' E_{out}/t_{decay}$, which assuming for simplicity $\eta'=1$ yields 
\begin{equation}
\label{eq:injection}
L'(t)=\frac{E_{out}}{t_{decay}} e^{-\frac{t-t_0}{t_{decay}}}.
\end{equation}

Taking into account that $E_{\Delta B}=V_{pwn} \Delta B^2 / (8 \pi)$, the evolution of the perturbation can be written as
\begin{equation}
\label{eq:deltabevol}
\frac{d(\Delta B)}{dt}=\frac{3 \eta'}{c^3} \frac{L'(t)}{\Delta B t^3}-\frac{2 \Delta B}{t},
\end{equation}
where $\Delta B$ is the magnetic field due to the additional injection; i.e., the total field being $B_{tot} = B_{pwn} + \Delta B$. Since in this phenomenological approach the additional field injected has a faster evolution in time than the original one residing in the nebula, the decay time of the PWN flux is shorter.

We assume that the X-ray emission coming from the PWN is synchrotron radiation. In a monochromatic approximation, the synchrotron power emitted by each electron (or positron) in the nebula is
\begin{equation}
P_{syn}(\nu,\gamma,t)=\frac{4}{3} \frac{\sigma_T}{m_e c} U_B \gamma^2 \delta(\nu-\nu_c),
\end{equation}
where $\nu$ is the frequency of the emitted photon, $\gamma$ the Lorentz factor of the electron, $t$ the time, $m_e$ the electron mass and $c$ the speed of light. The symbol $\sigma_T$ is the classical Thompson cross section for electrons, $U_B=B^2/(8 \pi)$ the magnetic energy density for a magnetic field $B$ and $\nu_c=3 e B \gamma^2/(4 \pi m_e c)$ is the so-called critical frequency where the maximum of power is emitted.  The synchrotron luminosity for an electron-positron distribution $N(\gamma,t)$ is then
\begin{equation}
L_{syn}(\nu,\gamma,t)=\int_{\gamma_{min}}^{\gamma_{max}} N(\gamma,t) P_{syn}(\nu,\gamma,t) \mathrm{d}\gamma.
\end{equation}
Note then that substituting $P_{syn}(\nu,\gamma,t)$ in the latter equation we have $L_{syn} \propto U_B \propto B^2$. Note that $B$ affects also the electron-positron population through synchrotron energy losses, but for a short timescale and low magnetic fields, we can consider $N(\gamma,t)$ approximately constant.  Then, if we want to detect an order of magnitude increase of the synchrotron luminosity $L_{syn,f}/L_{syn,i} \propto ((B+\Delta B)/B)^2 \simeq 10$, the change we should expect in the magnetic field is $\Delta B \simeq 2.16 B$.

The form of the energy injection and the evolution of $\Delta B$ are described in Eqs.~(\ref{eq:injection}) and~(\ref{eq:deltabevol}), respectively. The latter has an analytic solution, such that
\begin{equation}
\Delta B(t)=\frac{1}{(c t)^2} \sqrt{\frac{6 c E_{out}}{t_{decay}} \int_0^t t e^{-\frac{t}{t_{decay}}} \mathrm{d}t}
\end{equation}
where we set $t_0=0$ to simplify. This perturbation will affect the PWN when it reaches the termination shock. Most of the high energy particles are located close to that region, and then we will assume that once the termination shock is affected, all particles emitting in X-rays are affected by the increase of the magnetic field. At time $t_0$, the magnetic field perturbation starts to expand from the pulsar to the PWN. To estimate when this perturbation affects the PWN we need to gather the size of the termination shock by using the observational data available. The termination shock radius is given by (e.g. Gelfand et al. 2009).
\begin{equation}
\label{eq:ts}
R_{ts}(t)=\sqrt{\frac{\dot{E}}{4 \pi \chi c P_{pwn}}},
\end{equation}
where $\dot{E}$ is the spin-down luminosity of the pulsar, $P_{pwn}$ is the internal pressure of the PWN and $\chi$ is the filling factor of the wind which is assumed to be 1 for an isotropic wind. To calculate the internal pressure, we use the analytic function given in van der Swaluw et al. (2001):
\begin{equation}
\label{eq:pwnpressure}
P_{pwn}(t)=\frac{3}{25} \rho_{ej}(t) \left(\frac{R_{pwn}}{t} \right)^2,
\end{equation}
with $\rho_{ej}(t)$ is the density of the SNR ejecta defined as
\begin{equation}
\rho_{ej}(t)=\frac{3 M_{ej}}{4 \pi R_{pwn}^3(t)}
\end{equation}
being $M_{ej}$ the mass of the ejecta. We can recover $M_{ej}$ from Equation (4) and (12) in Van der Swaluw et al. (2001).
\begin{equation}
M_{ej}=7.04 \frac{E_{sn} t^2}{3 R_{pwn}^2} \left(\frac{\dot{E} t}{E_{sn}} \right)^{2/5}
\end{equation}
with $E_{sn}=10^{51}$ erg. Substituting the latter expression in Equation~\ref{eq:ts} and \ref{eq:pwnpressure}, we get an expression for the radius of the termination shock as a function of observational parameters
\begin{equation}
\label{eq:tsobs}
R_{ts}(t)=1.088 R_{pwn}(t)\sqrt{\frac{R_{pwn} \dot{E}}{c E_{sn}} \left(\frac{E_{sn}}{\dot{E} t} \right)^{2/5}}
\end{equation}

From the observations, we see that the radius of the PWN is $\sim$10$\arcsec$. Using the distance to J1119 as 8.4 kpc, we obtain $R_{pwn}\simeq$0.41 pc. The current spin-down luminosity is 2.3$\times$10$^{36}$ erg~s$^{-1}$ and its characteristic age 1.9 kyr. Taking into account all these data,  Eq.~\ref{eq:tsobs} yields $R_{ts} \simeq 0.0084$ pc, which means that the perturbation needs, at the speed of light, $\sim$10 days to reach the termination shock.

Due to the lack of data (outside the X-ray band) on the synchrotron spectrum of the PWN, we can not use a full radiative model to have an estimation of the magnetic field $B$ (e.g., Mart{\'\i}n et al. 2012), neither other rougher estimations, such as the ratio between the synchrotron and inverse Compton (IC) flux (see e.g., Aharonian et al. 1997). Thus, in order to get an estimation on the order of magnitude of $E_{out}$, we assume a mean magnetic field of 5 $\mu$G.  Such a low magnetic field can be justified by the fact that the X-ray emission is really dim in comparison to other PWNe, or to the spatially coincident TeV emission ($L_X/L_\gamma \sim 10^{-3}$, see Torres 2014).  Figure~4 shows the evolution of $\Delta B / B$ for $E_{out}=10^{41}$, $10^{42}$ and $10^{43}$ erg and $t_{decay}=0.1$, $0.5$, $1$, $2$ yr. The plot can be re-scaled for other values of $B$ such that $E_{out}=E_{out, 5\ \mu \text{G}} (B / 5\ \mu \text{G})^2$. Note that following our previous calculations, the ratio $\Delta B / B$ must be $\sim$2.16 when it crosses the vertical solid red line, this is the epoch of the observations reported in by Blumer et al. (2017). There is a continuum of values for $E_{out}$ and $t_{decay}$ that accomplish this condition, with $E_{out}$ around $10^{42}$~erg.

\begin{figure}[htp]
\centering
\includegraphics[width=.99\columnwidth]{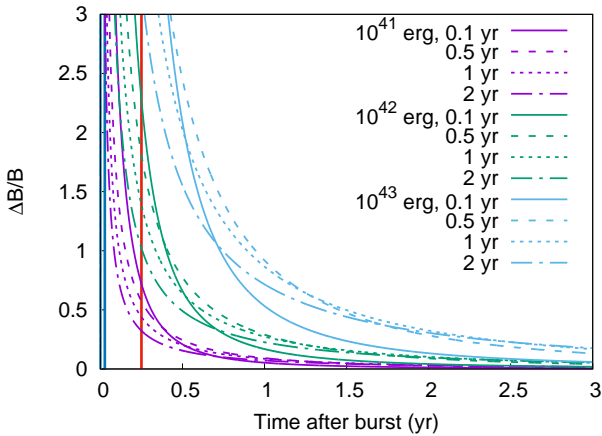}
\caption{Evolution of $\Delta B / B$ up to 3 yr after the outburst. The vertical solid blue line shows the moment when the perturbation reaches the termination shock. The vertical solid red line is the moment when the observations by Blumer et al. (2017) are done. The curves show the evolution for different values of $E_{out}$ and $t_{decay}$.}
\label{fig:deltabevol}
\end{figure}

We note that the total energetics needed to reproduce a PWN decay time of about three years, and an increase by a factor of 10 in the X-ray luminosity three months after the burst, is of course larger than the X-ray flux increase. In this model, the larger energetics is invested not only into producing the X-ray burst, but also local particles and magnetic field and perhaps the kinetic outflow itself, so that the X-ray measurement should function as a lower limit to the total power available. Among the several caveats of this simplified, but in principle working model, is that the decay time scale is governed by a damping process that is necessarily different from the one operating in the pre-burst PWN (and of which the equations above are just a phenomenological proxy).

\section{Conclusions}

In summary, we have conducted a deep Chandra observation of J1119 and its nebula, taken three years after the source went into outburst. Our results show that the pulsar has mostly returned to quiescence with its spectrum best fit by a combination of thermal and non-thermal components. A faint PWN is clearly detected, with jet-like features in the north-south direction, similar in morphology to the pre-burst phase. The PWN spectrum shows evidence of spectral softening in its quiescent post-outburst phase, with the pre-burst photon index $\Gamma$=1.2$\pm$0.4 changing to $\Gamma$=2.3$\pm$0.5. As well its pre-burst luminosity of 1.5$^{+0.5}_{-0.4}$$\times$10$^{32}$ erg~s$^{-1}$ is, three years following the outburst, is just slightly higher at 2.7$_{-0.2}^{+0.3}$$\times$10$^{32}$ erg~s$^{-1}$  (0.5--7 keV band). These changes, together with the observed timescale for returning to quiescence (of just a few years), imply a rather fast cooling process and favor a scenario where J1119 is temporarily powered by its magnetic energy following the magnetar-like outburst, in addition to its spin-down energy.

Further monitoring of J1119 and its PWN with Chandra would be beneficial to monitor the spectral evolution of this system. Obviously, should a new magnetar outburst happen in J1119, or in a different pulsar for which a known PWN exists, recurrent observations separated at months-scales would prove invaluable. That would allow us to track the increase and decrease of the nebula's luminosity and characterize its spectrum and morphology better to test the theoretical models.

\acknowledgements
This research has been supported by grants GO0-21047X (Chandra), PGC2018-095512-B-I00, SGR2017-1383 (Spain), and 2021VMA0001 (CAS). MAM is supported through NSF OIA award number 1458952. SSH acknowledges support by the Natural Sciences and Engineering Research Council of Canada (NSERC) and the Canadian Space Agency. AB is supported by a Juan de la Cierva Fellowship.

\software{CIAO (v14.12; Fruscione et al. 2006), XSPEC (v12.10.1f; Arnaud 1996), Sherpa (Freeman et al. 2001; Doe et al. 2007; Burke et al. 2020), ChaRT (Carter et al. 2003), MARX (Davis et al. 2012)}

\end{document}